\documentclass[pss]{wiley2sp} % provides new 2008 pss two-column style (no alternative manuscript style output available at present)
\usepackage{amsmath}
\usepackage{bm}              % uncomment these two packages if you
\usepackage{w-greek}         % need extended greek-letter functionality in math mode

\begin{document}

% Title of the article
\title{Reentrant phenomenon in the exactly solvable mixed spin-1/2 and spin-1 Ising-Heisenberg model
       on diamond-like decorated planar lattices}

% Abbreviated title for the page headers
\titlerunning{Reentrant phenomenon in the Ising-Heisenberg model on planar lattices}

% Authors
\author{%
  Lucia \v{C}anov\'a\textsuperscript{\textsf{\bfseries 1},\Ast},
  Jozef Stre\v{c}ka\textsuperscript{\textsf{\bfseries 2}}
       }

% Abbreviated list of authors for the page headers
\authorrunning{Lucia \v{C}anov\'a et al.}

%E-mail-address of corresponding author
\mail{e-mail
  \textsf{lucia.canova@tuke.sk}, Phone: +421 55 602 2228, Fax: +421 55 633 4738}

% author's affiliations/addresses
\institute{%
  \textsuperscript{1}\,Department of Applied Mathematics, Faculty of Mechanical Engineering,\\
Technical University, Letn\'a~9, 042~00 Ko\v{s}ice, Slovak Republic\\
  \textsuperscript{2}\,Department of Theoretical Physics and Astrophysics, Faculty of Science,\\
P.~J.~\v{S}af\'arik University, Park Angelinum~9, 040~01 Ko\v{s}ice, Slovak Republic
          }

\received{XXXX, revised XXXX, accepted XXXX} % do not change, will be filled in by the publisher
\published{XXXX} % do not change, will be filled in by the publisher

%Please select four to six PACS-codes from the enclosed list (PACS.txt) or from www.aip.org/pacs)
 % For example: 71.20.Ps

\abstract{%
\par{%
Ground-state and finite-temperature behaviour of the mixed spin-$1/2$ and spin-$1$ Ising-Heisenberg model
on decorated planar lattices consisting of inter-connected diamonds is investigated by means of the generalised decoration-iteration mapping transformation. The obtained exact results clearly point out that this model has a rather complex ground state composed of two unusual quantum phases, which is valid regardless of the lattice topology as well as the spatial dimensionality of the investigated system. It is shown that the diamond-like decorated planar lattices with a sufficiently high coordination number may exhibit a striking critical behaviour including reentrant phase transitions with two or three consecutive critical points.
    }
         }

\maketitle   % please do not remove

\section{Introduction}
Quantum Heisenberg models on {\it geometrically frustrated planar lattices} have enjoyed a great interest during the past decades especially due to their extraordinary diverse ground-state behaviour, which is often a result of mutual interplay between geometric frustration and quantum fluctuations~\cite{Lhu02,Rich04,Mis04}. From this perspective, geometrically frustrated quantum systems represent an excellent play ground for theoretical study of novel quantum many-body phenomena. Beside a rather complex ground state, which can be composed of several phases like the semi-classical N\'eel-like ordered phase, the quantum valence bond crystal phase or different disordered spin-liquid phases~\cite{Lhu02}, quantum spin systems on two-dimensional geometrically frustrated lattices furnish a deeper insight into the quantum order-from-disorder effect~\cite{Lhu02,Rich04,Mis04}, the scalar or vector chirality~\cite{Vil77,Vil80}, as well as, the non-zero residual entropy that characterizes the macroscopic degeneracy of the ground state~\cite{Lhu02}.

Among another interesting features, which also frequently attract the immense theoretical interest to the quantum Heisenberg model on geometrically frustrated planar lattices, belong an enhanced magnetocaloric effect emerging during the adiabatic demagnetisation~\cite{Zhi03,Hon05,Der06,Sch07} and a presence of quantized magnetisation plateaux in low-temperature magnetisation curves~\cite{Rich04,Hon04}. It is worthwile to remark that both these outstanding physical phenomena have also been observed in several real insulating magnetic materials representing possible experimental realizations of two-dimensional geometrically frustrated Heisenberg models. Recent experimental studies of the magnetocaloric effect in gadolinium galium garnet Gr$_3$Ga$_5$O$_{12}$~\cite{Bar82,Lac83} have revealed a potential applicability of frustrated quantum spin systems as promising refrigerant materials in the magnetic cooling technique
intended either for room temperature refrigeration or for satellite applications~(see~\cite{Zhi03} and references therein). The quantised magnetisation plateaux have been experimentally observed in the triangular lattice compounds ${\rm CsFe(SO_4)_2}$~\cite{Ina96}, ${\rm Cs_2CuBr_4}$~\cite{Tan02,Ono03,Ono04,Ono05a,Ono05b}, and ${\rm RbFe(MoO_4)_2}$~\cite{Ina96,Svi03,Pro03,Svi06,Smi07}, the kagom\'e lattice compound ${\rm[Cu_3(titmb)_2(CH_3COO)_6]\!\cdot\! H_2O}$~\cite{Nar04}, as well as the Shastry-Sutherland lattice compounds ${\rm SrCu_2(BO_3)_2}$~\cite{Kag99,Oni00,Kag01,Kag02}, $R{\rm B_4} (R = {\rm Er, Tm})$~\cite{Mich06,Yos06,Iga07,Sie08}.

Unfortunately, searching for the exact solution for the geometrically frustrated quantum Heisenberg models often fails due to a non-commutability between spin operators involved in their Hamiltonians. Owing to this fact, we have recently proposed a special class of geometrically frustrated {\it Ising-Heisenberg models on diamond-like decorated lattices}~\cite{Str02,Can04,Can06,Can08,Jas08,Str09,Can09}, which can be examined within the framework of an exact analytical approach based on the generalised decoration-iteration transformation~\cite{Fis59,Syo72,Roj09}. These simplified quantum models overcome the afore-mentioned mathematical difficulty by introducing the Ising spins at nodal lattice sites and the Heisenberg dimers on interstitial decorating sites of the considered planar lattice. It is worth mentioning that the mixed-spin Ising-Heisenberg models with diamond-like decorations turn out to be very useful testing ground for elucidating many quantum properties of low-dimensional magnetic materials, in spite of the fact that the monomer-dimer interaction is considered as the Ising-type interaction. Indeed, these rather simple spin models shed light on diverse quantum features to appear in the ground state~\cite{Str02,Can04,Can06,Can08,Can09}, the magnetisation process~\cite{Can06,Str09,Can09}, the thermodynamics~\cite{Can06,Str09}, as well as, the critical behaviour~\cite{Jas08}. Note that kinetically frustrated diamond chain models constituted by nodal Ising spins and mobile electrons delocalised over the interstitial decorating sites have been particularly examined as well~\cite{Per08,Per09,Lis}.

Motivated by these remarkable quantum features, being not commonly observed in any semi-classical Ising system, we will investigate in the present work the mixed spin-$1/2$ and spin-$1$ Ising-Heisenberg model on several decorated planar lattices consisting of inter-connected diamonds. Beside the ground-state analysis, our main goal is to particularly examine critical behaviour of these quantum models. We will also endeavour to explore how a possible existence of the reentrant phase transitions depends on a particular choice of the lattice topology of the investigated spin system.

The outline of this paper is as follows. In the next section, we will provide the detailed description of the Ising-Heisenberg model and then, some basic steps of the used decoration-iteration mapping method will be elucidated. Section~\ref{Sec3} deals with the most interesting results for the ground-state and finite-temperature phase diagrams, which are supplemented by a brief discussion of temperature variations of the total magnetisation and the specific heat. Finally, some concluding remarks are drawn in Section~\ref{Sec4}.

\section{Model and its exact solution \label{Sec2}}
\begin{figure*}[h]%
\vspace*{-0.75cm}
\hspace*{2.75cm}
\includegraphics*[width=.825\textwidth]{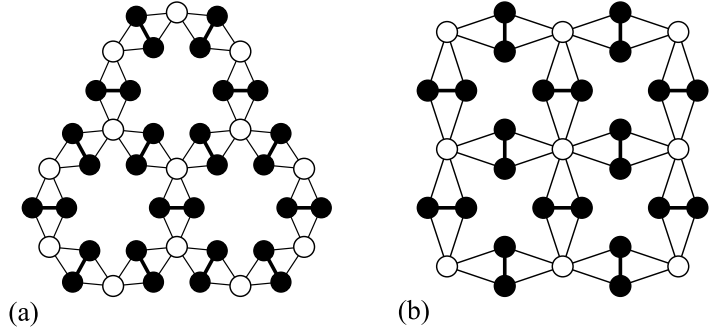}%
\vspace*{-1.5cm}
\caption{The mixed spin-$1/2$ and spin-$1$ Ising-Heisenberg model on diamond-like decorated honeycomb
         [Fig.~\ref{fig1}(a)] and square [Fig.~\ref{fig1}(b)] lattices. White circles represent the nodal Ising spins
         $\sigma = 1/2$ and black ones denote the decorating Heisenberg spins $S = 1$.}
\label{fig1}
\end{figure*}

Let us consider two-dimensional lattices composed of inter-connected diamonds as is illustrated in Fig.~\ref{fig1} for honeycomb and square lattices.
\begin{figure}[htb]%
\vspace*{-1.0cm}
\includegraphics*[width=\linewidth]{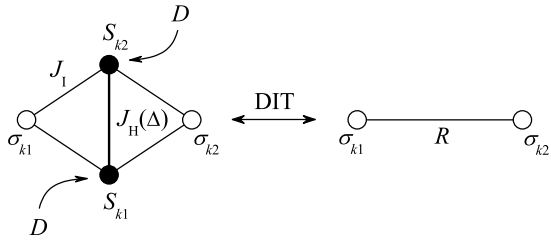}%
\vspace*{-1.0cm}
\caption{A schematic representation of the decoration-iteration transformation of the kth diamond-shaped unit of
         the investigated mixed-spin Ising-Heisenberg model (lhs of figure) towards the unit of the simple spin-1/2
         Ising model on corresponding lattice with the nearest-neighbour interaction $R$ (rhs of figure). The parameters
         $J_{\rm I}$, $J_{\rm H}(\Delta)$, and $D$ stand for the exchange interaction between the nearest-neigbouring
         Heisenberg spins, the interaction between the nearest-neigbouring Ising and Heisenberg spins and the axial
         zero-field splitting parameter, respectively.}
\label{fig2}
\end{figure}
As shown in this figure, the diamond-shaped unit of considered lattices consists of two different kinds of spin sites schematically depicted as empty and full circles. To ensure an exact tractability of the investigated spin system, we will further suppose that the former spin sites are occupied by the {\it Ising spins} $\sigma = 1/2$ that interact with other spins through the interaction $J_{\rm I}$ containing just one ($z$th) spatial component of the spin operator, while the latter ones are occupied by the decorating {\it Heisenberg spins} $S = 1$ that interact among themselves via the anisotropic XXZ coupling $J_{\rm H}(\Delta)$ containing all three spatial components of spin operators. Note that the parameter $\Delta$ allows to control the interaction $J_{\rm H}$ between the easy-axis ($\Delta < 1$) and easy-plane ($\Delta > 1$) type, as well as, to obtain the Ising model as a special limiting case when assuming $\Delta = 0$. Under these assumptions, the total Hamiltonian of the investigated model can be written as
\begin{eqnarray}
\hat{{\cal H}} =
&-&\! J_{\rm H}\sum_{(i,j)}^{Nq/2}\,
[\Delta(\hat{S}_{i}^{x}\hat{S}_{j}^{x}+\hat{S}_{i}^{y}\hat{S}_{j}^{y})+\hat{S}_{i}^{z}\hat{S}_{j}^{z}]
\nonumber \\
&-&\! J_{\rm I}\sum_{(i,n)}^{2Nq} \hat{S}_{i}^{z}\hat{\sigma}_{n}^{z}
 - D\sum_{i = 1}^{2Nq}\, (\hat{S}_{i}^{z})^{2}.
\label{eq:Hih}
\end{eqnarray}
Here, the first summation is carried out over all exchange interactions between pairs of the nearest-neigbouring Heisenberg spins, the second summation takes into account the interaction between the nearest-neigbouring Ising and Heisenberg spins and the last summation runs over all lattice sites occupied by the decorating Heisenberg spins. The spin variables $\hat{S}_{i}^{\gamma}$ ($\gamma = x, y, z$) and $\hat{\sigma}_{n}^{z}$ denote spatial components of the spin-$1/2$ and spin-$1$ operators located at $i$th and $n$th lattice position, respectively. The parameter $D$ stands for the axial zero-field splitting (AZFS) parameter acting on the Heisenberg spins only~\cite{Rud08a,Rud08b}. Finally, $N$ represents the total number of the Ising spins and $q$ is the coordination number of the lattice, which implies that the total number of all spins (lattice sites) is $N_{tot} = N(q+1)$.

The crucial step of our calculation represents the evaluation of the partition function of the system. In view of further manipulations, it is useful to rewrite the total Hamiltonian~(\ref{eq:Hih}) as a sum over the cluster Hamiltonians $\hat{{\cal H}} = \sum_{k = 1}^{Nq/2} \hat{{\cal{H}}}_k$, where each cluster Hamiltonian $\hat{{\cal{H}}}_k$ involves all interaction terms that belong to just one ($k$th) diamond unit (see lhs of Fig.~\ref{fig2}):
\begin{eqnarray}
&&\hat{{\cal H}}_k =
- J_{\rm H}[\Delta(\hat{S}_{k1}^{x}\hat{S}_{k2}^{x}+\hat{S}_{k1}^{y}\hat{S}_{k2}^{y})+\hat{S}_{k1}^{z}\hat{S}_{k2}^{z}]
\nonumber \\
&&- J_{\rm I}(\hat{S}_{k1}^{z} + \hat{S}_{k2}^{z})(\hat{\sigma}_{k1}^{z} + \hat{\sigma}_{k2}^{z})
- D[(\hat{S}_{k1}^{z})^{2} + (\hat{S}_{k2}^{z})^{2}].
\label{eq:Hk}
\end{eqnarray}
Taking into account a validity of the commutation relation between cluster Hamiltonians of two different diamond units $[\hat{\mathcal{H}}_k, \hat{\mathcal{H}}_l] = 0$, the partition function of the investigated mixed-spin Ising-Heisenberg model can be partially factorised into the following product of cluster partition functions ${\cal Z}_k$:
\begin{equation}
{\cal Z} = \sum_{\{ \sigma_i \}} \prod_{k=1}^{Nq/2} {\rm Tr}_{k} \exp(-\beta\hat{{\mathcal H}_k}) = \sum_{\{ \sigma_i \}} \prod_{k=1}^{Nq/2} {\cal Z}_{k}.
\label{eq:Z}
\end{equation}
Here, $\beta = 1/(k_{B}T)$, $k_{\rm B}$ is Boltzmann's constant and $T$ is the absolute temperature. The summation $\sum_{\{ \sigma_i \}}$ in Eq.~(\ref{eq:Z}) runs over all possible spin configurations of the nodal Ising spins, while the symbol ${\rm Tr}_{k}$ stands for a trace over degrees of freedom of the $k$th Heisenberg dimer. By performing an exact analytical diagonalisation of the cluster Hamiltonian (\ref{eq:Hk}) in a particular Hilbert subspace corresponding to the Heisenberg spin pair from $k$th diamond unit, the partition function ${\cal Z}_{k}$ will depend on two nodal Ising spins $\sigma_{k1}$ and $\sigma_{k2}$ only. Moreover, the resulting explicit form of ${\cal Z}_{k}$ immediately implies a possibility of performing the generalised decoration-iteration mapping transformation~\cite{Fis59,Syo72,Roj09}
\begin{eqnarray}
{\cal Z}_{k} &=&  {\rm e}^{2\beta D - \beta J_{\rm H}} + 2\,{\rm e}^{2\beta D + \beta J_{\rm H}}\cosh[2\beta J_{\rm I}(\sigma_{k1}^z \!+ \sigma_{k2}^z)]
\nonumber \\
&+& 2\,{\rm e}^{\beta D - \beta J_{\rm H}/2}\cosh[\beta\sqrt{(2D - J_{\rm H})^2\! + 8(J_{\rm H}\Delta)^2}/2]
\nonumber \\
&+& 4\,{\rm e}^{\beta D}\cosh[\beta J_{\rm I}(\sigma_{k1}^z \!+ \sigma_{k2}^z)]\cosh(\beta J_{\rm H}\Delta)
\nonumber \\
&=& A\, {\rm e}^{\beta R \sigma_{k1}^z\sigma_{k2}^z}.
\label{eq:DIT}
\end{eqnarray}
From the physical point of view, the mapping transformation~(\ref{eq:DIT}) removes all interaction parameters associated with the Heisenberg spins and replaces them by the effective interaction $R$ between the remaining nodal Ising spins $\sigma_{k1}$ and $\sigma_{k2}$ (see Fig.~\ref{fig2} for better illustration). Of course, the transformation relation~(\ref{eq:DIT}) must hold for any available spin configuration of the Ising spins. This ''self-consistency'' condition unambiguously determines so far not specified mapping parameters $A$ and $R$,
\begin{equation}
A = \sqrt{W_1 W_2}\,, \quad
R = 2\beta^{-1}\!\left(\ln W_1 - \ln W_2\right).
\label{eq:AR}
\end{equation}
It is clear that the functions $W_1$ and $W_2$ emerging in the mapping parameters (\ref{eq:AR}) are two independent expressions to be obtained by substituting four possible configurations of the Ising spins into the cluster partition function~(\ref{eq:DIT}):
\begin{eqnarray}
W_1 &=& {\rm e}^{2\beta D - \beta J_{\rm H}} + 2\,{\rm e}^{2\beta D +\beta J_{\rm H}}\cosh(2\beta J_{\rm I})
\nonumber \\
&+& 2\,{\rm e}^{\beta D - \beta J_{\rm H}/2}\cosh[\beta \sqrt{(2D - J_{\rm H})^2 + 8(J_{\rm H}\Delta)^2}/2]
\nonumber \\
&+& 4\,{\rm e}^{\beta D} \cosh(\beta J_{\rm I})\cosh(\beta J_{\rm H}\Delta),
\\
W_2 &=& {\rm e}^{2\beta D - \beta J_{\rm H}} + 2\,{\rm e}^{2\beta D +\beta J_{\rm H}}
\nonumber \\
&+& 2\,{\rm e}^{\beta D - \beta J_{\rm H}/2}\cosh[\beta \sqrt{(2D - J_{\rm H})^2 + 8(J_{\rm H}\Delta)^2}/2]
\nonumber \\
&+& 4\,{\rm e}^{\beta D}\cosh(\beta J_{\rm H}\Delta).
\label{eq:W1W2}
\end{eqnarray}
At this stage, the straightforward substitution of the transformation relation~({\ref{eq:DIT}}) into the expression~({\ref{eq:Z}}) yields the equality
\begin{equation}
{\cal Z} (T, J_{\rm I} , J_{\rm H}, \Delta, D) = A^{Nq/2} {\cal Z}_{\rm Ising}(T, R),
\label{eq:Zih=Zi}
\end{equation}
which establishes an exact mapping relationship between the partition function ${\cal Z}$ of the mixed spin-$1/2$ and spin-$1$ Ising-Heisenberg model on diamond-like decorated planar lattices and the partition function ${\cal Z}_{\rm Ising}$ of the simple spin-$1/2$ Ising model on corresponding undecorated lattices with the nearest-neighbour coupling $R$. In this respect, Eq.~(\ref{eq:Zih=Zi}) formally completes our calculation for the partition function of the spin-$1/2$ and spin-$1$ Ising-Heisenberg model, since exact results for partition functions of many spin-$1/2$ Ising planar lattices are known~\cite{Syo72}. It is worth noticing that this mapping relation is universal and valid regardless of the lattice topology and space dimensionality of the investigated model system. Besides, it also allows a direct calculation of some physical quantities (such as Gibbs free energy, internal energy, magnetisation, entropy, susceptibility, or specific heat) by the use of standard thermodynamical-statistical relations, which are important for understanding of the magnetic behaviour of the investigated spin system. Other important physical quantities, such as sub-lattice magnetisation~$m_{\rm i}^z$ and~$m_{\rm h}^z$ reduced per one Ising and Heisenberg spin, respectively, correlation functions~$c_{\rm ii}^{zz}$, $c_{\rm hh}^{zz}$, $c_{\rm hh}^{xx}$, and $c_{\rm ih}^{zz}$, as well as, quadrupolar moment~$q_{\rm hh}^{zz}$, which cannot be obtained within afore-mentioned procedure, can be calculated by combining the relation~(\ref{eq:Zih=Zi}) with the exact mapping theorems developed by Barry {\it et al.}~\cite{Bar88,Kha90,Bar91} and the generalized Callen-Suzuki spin identity~\cite{Cal63,Suz65,Bal02}:
\begin{eqnarray}
\label{eq:miz}
m_{\rm i}^z &\equiv& \langle \hat{\sigma}_{k1}^{z}
\rangle = \langle \hat{\sigma}_{k1}^{z} \rangle_{0} \equiv m_0,
\\
\label{eq:mhz}
m_{\rm h}^z &\equiv& \langle \hat{S}_{k1}^{z}\rangle =
4m_0 f_1(J_{\rm I}),
\\
\label{eq:ciz}
c_{\rm ii}^{zz} &\equiv& \langle \hat{\sigma}_{k1}^{z}
\hat{\sigma}_{k2}^{z} \rangle = \langle \hat{\sigma}_{k1}^{z}
\hat{\sigma}_{k2}^{z} \rangle_0 \equiv \varepsilon_0,
\\
\label{eq:chz}
c_{\rm hh}^{zz} &\equiv& \langle \hat{S}_{k1}^{z}
\hat{S}_{k2}^{z} \rangle = \left(1 \!+ 4\varepsilon_0\right)\!f_2(\!J_{\rm I}) + \left(1 \!- 4\varepsilon_0\right)\!f_2(0),
\\
\label{eq:chx}
c_{\rm hh}^{xx} &\equiv& \langle \hat{S}_{k1}^{x}
\hat{S}_{k2}^{x} \rangle = \left(1 \!+ 4\varepsilon_0\right)\!f_3(\!J_{\rm I}) + \left(1 \!- 4\varepsilon_0\right)\!f_3(0),
\\
\label{eq:cihz}
c_{\rm ih}^{zz} &\equiv& \langle \hat{\sigma}_{k1}^{z}
\hat{S}_{k1}^{z} \rangle = \left(1/2 + 2\varepsilon_0\right)\!f_1(\!J_{\rm I}),
\\
\label{eq:qhz}
q_{\rm hh}^{zz} &\equiv& \langle (\hat{S}_{k1}^{z})^2
\rangle = \left(1 \!+ 4\varepsilon_0\right)\!f_4(\!J_{\rm I}) + \left(1 \!- 4\varepsilon_0\right)\!f_4(0).
\end{eqnarray}
In above, the symbols $\langle\ldots\rangle$ and $\langle\ldots\rangle_0$ represent standard canonical averages performed over the ensemble defined by the mixed-spin Ising-Heisenberg model on the diamond-like decorated lattice and the spin-$1/2$ Ising model on the corresponding lattice, respectively, $m_0$ labels the single-site magnetisation and $\varepsilon_0$~stands for the nearest-neighbour correlation function of the corresponding Ising model. Finally, the functions $f_i(x)$ ($i = 1$--$4$), emerging in the set of Eqs.~(\ref{eq:mhz}), (\ref{eq:chz})--(\ref{eq:qhz}), are defined as follows:
\begin{eqnarray}
f_1(x) &=&
[{\rm e}^{\beta D +\beta J_{\rm H}} \sinh(2\beta x) {}
\nonumber\\
&+& \sinh(\beta x)\cosh(\beta J_{\rm H}\Delta)]/G(x),
\\
f_2(x) &=& [2\,{\rm e}^{\beta D +\beta J_{\rm H}}\cosh(2\beta x) - {\rm e}^{\beta D - \beta J_{\rm H}} {}
\nonumber\\
&-& (2D -\! J_{\rm H})\omega^{-1}{\rm e}^{-\beta J_{\rm H}/2} \sinh(\beta\omega/2) {}
\nonumber\\
&-&\, {\rm e}^{-\beta J_{\rm H}/2} \cosh(\beta \omega/2)]/[2G(x)],
\\
f_3(x) &=& [\cosh(\beta x)\sinh(\beta J_{\rm H}\Delta) {}
\nonumber\\
&+& 2J_{\rm H}\Delta\omega^{-1}{\rm e}^{-\beta J_{\rm H}/2}\sinh(\beta \omega/2)]/G(x),
\end{eqnarray}
\begin{eqnarray}
f_4(x) &=& [2\,{\rm e}^{\beta D +\beta J_{\rm H}}\cosh(2\beta x) + {\rm e}^{-\beta J_{\rm H}/2} \cosh(\beta\omega/2) {}
\nonumber \\
&+& (2D -\! J_{\rm H})\omega^{-1}{\rm e}^{-\beta J_{\rm H}/2}
\sinh(\beta \omega/2) + {\rm e}^{\beta D - \beta J_{\rm H}}
\nonumber \\
&+& 2\cosh(\beta x)\cosh(\beta J_{\rm H}\Delta)   ]/[2G(x)],
\end{eqnarray}
where $G(x)={\rm e}^{\beta D - \beta J_{\rm H}} + 2\,{\rm e}^{\beta D +\beta J_{\rm H}}\cosh(2\beta x)
+ 4\cosh(\beta x)\cosh(\beta J_{\rm H}\Delta) + 2\,{\rm e}^{-\beta J_{\rm H}/2}\cosh(\beta\omega/2)$ and $\omega = \sqrt{(2D - J_{\rm H})^2 + 8(J_{\rm H}\Delta)^2}$.

\section{Results and discussion \label{Sec3}}
Before proceeding to a discussion of the most interesting results obtained for the mixed spin-$1/2$ and spin-$1$ Ising-Heisenberg model on diamond-like decorated honeycomb and square lattices, let us make a few remarks on a validity of analytical results presented in the previous section. It is worth mentioning that all obtained results are universal as they hold regardless of whether ferromagnetic or antiferromagnetic interaction parameters $J_{\rm I}$ and $J_{\rm H}$ are assumed, as well as, independently of the lattice topology or spatial dimensionality of the investigated spin system. As proved, however, there are some fundamental differences between magnetic behaviour of models with distinct nature of the Heisenberg interaction (see our preliminary reports~\cite{Str02,Can08,Jas08}). Considering this fact, we will restrict ourselves here just to the case with the ferromagnetic Heisenberg interaction $J_{\rm H}>0$. To simplify our discussion, we will also assume the Ising interaction $J_{\rm I}$ to be ferromagnetic, because the sign change $J_{\rm I} \rightarrow - J_{\rm I}$ brings just a trivial change in the local alignment of the nodal Ising spins with respect to their nearest Heisenberg neighbours. Indeed, it changes the parallel alignment between the Ising and Heisenberg spins along the $z$-axis to the antiferromagnetic one. Other particular case of the system, in which both the interaction constants $J_{\rm H}$ and $J_{\rm I}$ are supposed to be antiferromagnetic ($J_{\rm H}<0$, $J_{\rm I}<0$), will be explored in the subsequent work~\cite{Str}. Bearing this in mind, we can set the Ising exchange interaction $J_{\rm I}$ as the energy unit and introduce the following set of dimensionless parameters: $t = k_{\rm B}T/J_{\rm I}$, $d = D/J_{\rm I}$, and~$\alpha = J_{\rm H}/J_{\rm I}$, as describing the dimensionless temperature, the relative strength of the AZFS parameter, and the strength of the Heisenberg interaction normalized with respect to the Ising interaction, respectively.

\subsection{Ground-state properties}
Let us start our discussion with the analysis of the ground-state phase diagrams displayed in Fig.~\ref{fig3}. As one can see from the phase diagram constructed in the $\Delta - \ln\alpha$ plane [Fig.~\ref{fig3}(a)], three different phases appear in total within the ground state of the system without AZFS parameter due to a mutual interplay between the exchange interactions $J_{\rm I}$, $J_{\rm H}$, and the exchange anisotropy $\Delta$. It is noteworthy that a presence of non-zero AZFS paramer does not lead to any new phase as illustrated on the ground-state phase diagram from Fig.~\ref{fig3}(b). Spin arrangements of three possible ground-state phases can be unambiguously characterised as follows:
\begin{figure}[htb]%
\vspace*{-0.5cm}
\includegraphics*[width=7.5cm]{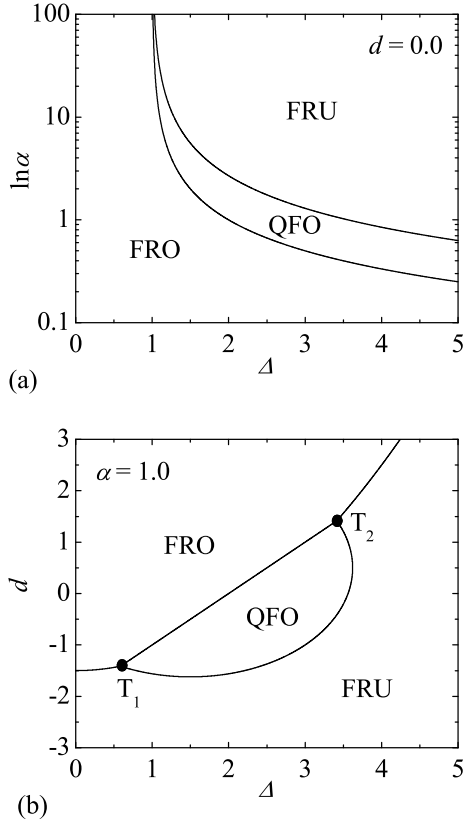}%
\vspace*{-0.5cm}
\caption{Ground-state phase diagrams of the mixed spin-$1/2$ and spin-$1$ Ising-Heisenberg model with ferromagnetic
         Ising and Heisenberg exchange interactions (a)~in the $\Delta - \ln\alpha$ plane for the system without
         the AZFS parameter and (b)~in the $\Delta - d$ plane for the system with $\alpha = 1.0$. Spin orders to
         emerge within the phase sectors FRO, QFO, and FRU are unambiguously determined by the
         wave-functions~(\ref{eq:FRO})--(\ref{eq:FRU}).}
\label{fig3}
\end{figure}
\begin{eqnarray}
|{\rm FRO} \rangle &=&
\prod_{i=1}^{N} | + \rangle_{i}\!
\prod_{k=1}^{Nq/2} | 1, 1  \rangle_{k}, \nonumber \\
&&\hspace{-1.33cm} m_{\rm i}^z = 0.5, m_{\rm h}^z = 1, c_{\rm ii}^{zz} = 0.25, \nonumber \\
&&\hspace{-1.33cm} q_{\rm hh}^{zz} = 1, c_{\rm hh}^{zz} = 1, c_{\rm hh}^{xx} = 0, c_{\rm ih}^{zz} = 0.5;
\label{eq:FRO}
\\[0.2cm]
|{\rm QFO} \rangle &=&
\prod_{i=1}^{N} | + \rangle_{i}\!
\prod_{k=1}^{Nq/2} \frac{1}{\sqrt{2}} \Bigl(|1,0 \rangle_{k} + |0,1 \rangle_{k} \Bigr),
\nonumber \\
&&\hspace{-1.33cm} m_{\rm i}^z = 0.5, m_{\rm h}^z = 0.5, c_{\rm ii}^{zz} = 0.25, \nonumber \\
&&\hspace{-1.33cm} q_{\rm hh}^{zz} = 0.5, c_{\rm hh}^{zz} = 0, c_{\rm hh}^{xx} = 0.5, c_{\rm ih}^{zz} = 0.25;
\label{eq:QFO}
\\[0.2cm]
|{\rm FRU} \rangle &=&
\prod_{i=1}^{N} |\pm \rangle_{i}\!
\prod_{k=1}^{Nq/2} \frac{1}{2} \Bigl[ b_{+} \Bigl( |1,-1  \rangle_{k} + |-1,1  \rangle_{k} \Bigr) \nonumber \\
&&\hspace{2.33cm}                     + \sqrt{2}b_{-} |0, 0 \rangle_{k} \Bigr],
\nonumber \\
&&\hspace{-1.33cm} m_{\rm i}^z = 0, m_{\rm h}^z = 0, c_{\rm ii}^{zz} = 0,\nonumber \\
&&\hspace{-1.33cm} q_{\rm hh}^{zz} = - c_{\rm hh}^{zz} = b_{+}^{2}/2, c_{\rm hh}^{xx} = 2J_{\rm H}\Delta\omega^{-1}, c_{\rm ih}^{zz} = 0,
\label{eq:FRU}
\end{eqnarray}
where $b_{\pm} =  \sqrt{1 \pm (2D - J_{\rm H})\omega^{-1}}$. In above, the first product is taken over all nodal Ising spins, while the second one runs over all Heisenberg dimers. Ket vectors $|\pm \rangle$ and $|\pm 1, 0 \rangle$ after relevant product symbols determine the spin states $\sigma^z = \pm 1/2$ and $S^z = \pm 1, 0$ of the Ising and Heisenberg spins, respectively. Analytic expressions for the phase boundaries between these ground states read:
\begin{eqnarray}
&&|{\rm FRO} \rangle - |{\rm QFO} \rangle: \,\, d = \alpha(\Delta - 1) - 1,
\\
&&|{\rm FRO} \rangle - |{\rm FRU} \rangle: \,\, d = -\frac{\alpha}{2} - 1 + \frac{(\alpha\Delta)^2}{2(\alpha+1)},
\\
&&|{\rm QFO} \rangle - |{\rm FRU} \rangle: \,\, \Bigl[\alpha\Bigl(\Delta - \frac{1}{2}\Bigr) - 1\Bigr]^2 \!+ \Bigl(d - \frac{\alpha}{2}\Bigr)^2 
\nonumber \\
&& \hspace{3.cm}= 2\Bigl(1+\frac{\alpha}{2}\Bigr)^2,
\end{eqnarray}
which all meet at two triple points given by the condition
\begin{eqnarray}
{\rm T}_{1,2} = [\Delta_{\rm T}, d_{\rm T}] = \left[\frac{1+\alpha}{\alpha} \mp \frac{\sqrt{1+\alpha}}{\alpha}, \mp\frac{\sqrt{1+\alpha}}{\alpha}\right]\!.
\end{eqnarray}
As one clearly sees from Eqs.~(\ref{eq:FRO}), the phase FRO represents the standard ferromagnetic phase, which can commonly be observed in the pure Ising systems as well. Actually, all the results indicate a perfect parallel alignment of the nearest-neighbouring Ising and Heisenberg spins in this phase. By contrast, other two phases QFO and FRU exhibit very unusual spin arrangements that cannot be observed in any semi-classical Ising systems. More specifically, all nodal Ising spins are perfectly aligned along the $z$-axis, while the decorating Heisenberg ones reside at a quantum superposition of spin states described by the symmetric wave function $(|1,0 \rangle + |0,1 \rangle)/\sqrt{2}$ in the QFO phase. Moreover, one may conclude from the location of the QFO phase in the phase diagram shown in Fig.~\ref{fig3}(a) that this phase appears as a result of a mutual competition between the Ising interaction $J_{\rm I}$ favouring the ferromagnetic spin arrangement along the $z$-axis, and the easy-plane Heisenberg interaction $J_{\rm H}(\Delta)$, which energetically favours the short-range ferromagnetic spin order in the $xy$ plane. Of course, the position of QFO changes as the AZFS parameter $d$ is switched on: for $d > 0$, one finds a lower range of values of the exchange anisotropy $\Delta$ upon strengthening the AZFS parameter $d$ corresponding to this phase, whereas this range at first extends and only then shrinks when $d < 0$ [see Fig.~\ref{fig3}(b)]. Another phase FRU also appears as a result of the mutual competition between the exchange interactions $J_{\rm I}$ and~$J_{\rm H}(\Delta)$ with $\Delta > 1$, respectively. As one can readily deduce from Eqs.~(\ref{eq:FRU}), this phase represents a macroscopically degenerate monomer-dimer state, where all nodal Ising spins are frustrated due to antiferromagnetic and/or 'non-magnetic' nature of the Heisenberg spin dimers. 
\begin{figure}[htb]%
\vspace*{-0.5cm}
\includegraphics*[width=\linewidth]{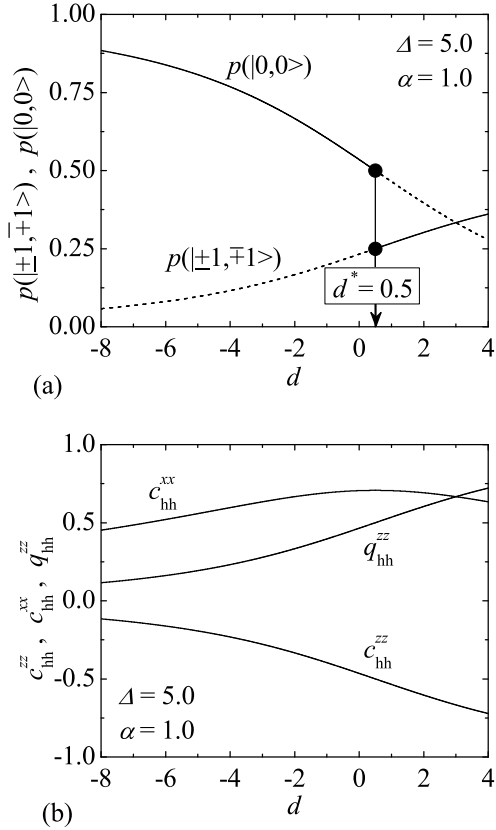}%
\vspace*{-0.5cm}
\caption{(a)~The probability distribution for the entangled spin states of the Heisenberg spin pairs in the FRU phase
         as a function of the AZFS parameter $d$ for $\Delta = 5.0$ and $\alpha = 1.0$.
         (b)~The corresponding zero-temperature dependencies of the quadrupolar moment~$q_{\rm hh}^{zz}$
         and correlation functions~$c_{\rm hh}^{zz}$, $c_{\rm hh}^{xx}$. For clarity, the spin state with the major
         (minor) abundance is depicted  in Fig.~\ref{fig4}(a) by full (broken) line.}
\label{fig4}
\end{figure}
Actually, the Heisenberg spin pairs exhibit a quantum superposition of three spin states $|0, 0\rangle$, $|1,-1\rangle$, and $|-1,1\rangle$ in the FRU phase, whose relative probabilities depend on a mutual ratio between the interaction parameters $J_{\rm H}$, $\Delta$, and $D$ [see the eigenfunction in~(\ref{eq:FRU})]:
\begin{eqnarray}
&&p(|\pm 1,\mp 1\rangle) =
\frac{1}{4}\!\left(1 + \frac{2d - \alpha}{\sqrt{(2d-\alpha)^2+8(\alpha\Delta)^2}}\right)\!, \\
&&p(| 0, 0\rangle) =
\frac{1}{2}\!\left(1 - \frac{2d - \alpha}{\sqrt{(2d-\alpha)^2+8(\alpha\Delta)^2}}\right)\!.
\label{eq:pp}
\end{eqnarray}
If $2p(|\pm 1,\mp 1\rangle) < p(| 0, 0\rangle)$, then the most probable spin state of the Heisenberg spin pairs is the 'non-magnetic' state $|0, 0\rangle$, while the antiferromagnetic $|\pm 1,\mp 1 \rangle$ spin states become major microstates if $2p(|\pm 1,\mp 1\rangle) > p(| 0, 0\rangle)$. According to this condition, the threshold value of the AZFS parameter dividing the parameter space of the FRU phase into a region with the dominant population of the 'non-magnetic' $|0, 0\rangle$ spin state and a region with the prevailing population of the antiferromagnetic $|\pm 1,\mp 1 \rangle$ spin states is $d^* = 0.5\alpha$.

More obvious insight into a probability distribution of both afore-mentioned kinds of microstates to emerge in the FRU phase may also be gained from Figs.~\ref{fig4}(a) and~(b), which show the abundance probabilities for relevant spin states as functions of the AZFS parameter $d$ for $\Delta = 5.0$ and $\alpha = 1.0$, as well as, the corresponding zero-temperature dependencies of the quadrupolar moment~$q_{\rm hh}^{zz}$ and the correlation functions~$c_{\rm hh}^{zz}$, $c_{\rm hh}^{xx}$, respectively. Note that the curves corresponding to the probabilities $p(|\pm 1,\mp 1\rangle)$ and $p(| 0, 0\rangle)$ are for clarity depicted in Fig.~\ref{fig4}(a) as full (broken) lines when the spin configuration $|\pm 1,\mp 1\rangle$ and $| 0, 0\rangle$ of the Heisenberg spin pairs has the major (minor) abundance in the FRU phase. As one can see from this figure, the probability to find the Heisenberg spin pairs in the 'non-magnetic' $|0, 0\rangle$ spin state gradually decreases upon increase of the AZFS parameter, while the probability of finding two antiferromagnetic $|\pm 1,\mp 1 \rangle$ spin states gradually increases with increasing the AZFS parameter $d$. As a result, the abundance of these two kinds of spin states changes as soon as the threshold value of the AZFS parameter $d^*$ is attained. Above this boundary value, the population of the antiferromagnetic $|\pm 1,\mp 1 \rangle$ ('non-magnetic' $|0, 0\rangle$) spin states becomes major (minor) and the probability of its finding further increases (decreases) as $d$ increases.

As an independent check of this scenario, we have plotted the zero-temperature dependencies of the pair correlation function~$c_{\rm hh}^{zz}$ and the quadrupolar moment~$q_{\rm hh}^{zz}$ in Fig.~\ref{fig4}(b). Obviously, both these physical quantities as functions of the parameter~$d$ exhibit monotonous dependencies: the correlation function~$c_{\rm hh}^{zz}$ monotonically decreases, while the quadrupolar moment~$q_{\rm hh}^{zz}$ monotonically increases upon increase of $d$. By contrast, the correlation function~$c_{\rm hh}^{xx}$ exhibits a much more complex variation: it first increases to reach its maximum at $d^*$ and then decreases with the increasing $d$. As expected, this non-monotonic behaviour of~$c_{\rm hh}^{xx}$ relates to a different trend of the short-range order of the Heisenberg spins in the region $d < d^*$ with the most populated 'non-magnetic' $|0, 0\rangle$ spin state and the region $d > d^*$ with the most populated antiferromagnetic $|\pm 1,\mp 1 \rangle$ spin states. In the former region, the increase in the AZFS parameter $d$ reinforces the influence of the easy-plane exchange anisotropy $\Delta > 1$. In the consequence of that, the short-range antiferromagnetic order along the $z$-axis is accompanied with the short-range ferromagnetic order in the $xy$ plane. The precisely opposite situation emerges in the latter region: for any $d < d^*$ the exchange anisotropy $\Delta > 1$ is restrained by the parameter $d$, which leads to the destruction of the short-range ferromagnetic order of the Heisenberg spins in the $xy$ plane on behalf of	short-range antiferromagnetic spin order along the $z$-axis whenever $d$ increases.

\subsection{Finite-temperature behaviour}
\begin{figure*}[htb]%
\vspace*{-0.75cm}
\begin{center}
\includegraphics*[width=14.5cm]{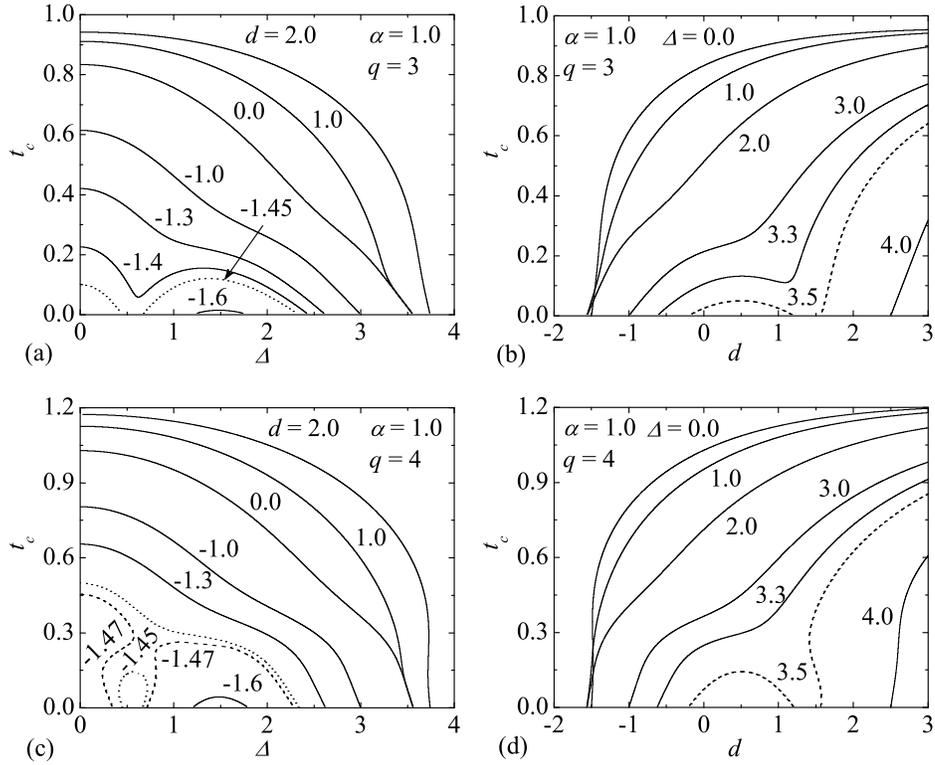}
\end{center}
\vspace*{-0.75cm}
\caption{The critical temperature of the mixed spin-$1/2$ and spin-$1$ Ising-Heisenberg model on diamond-like decorated
         honeycomb [Figs.~\ref{fig5}(a), (b)] and square [Figs.~\ref{fig5}(c), (d)] lattices at the fixed interaction
         ratio~$\alpha = 1.0$ as a function of the exchange anisotropy~$\Delta$ for several values of the AZFS
         parameter $d$ and as a function of the AZFS parameter $d$ for several values of the exchange
         anisotropy~$\Delta$.}
\label{fig5}
\end{figure*}
To provide a deeper insight into the finite-temperature behaviour of the investigated model system, let us turn our attention to the phase diagrams displayed in Fig.~\ref{fig5}. This figure shows the critical temperature of the mixed spin-$1/2$ and spin-$1$ Ising-Heisenberg model on diamond-like decorated honeycomb and square lattices as a function of the exchange anisotropy~$\Delta$ and the AZFS parameter $d$ for the fixed interaction ratio~$\alpha = 1.0$.
As it can be clearly seen, the model under investigation displays a rather complex critical behaviour that very sensitively depends on a strength of both parameters~$\Delta$ and $d$, as well as, the topology (coordination number) of the investigated lattice. Apart from the expected decrease of $t_c$ with the increasing $\Delta$ and/or decreasing $d$, one may also find non-monotonic dependencies of $t_c$ to be closely related to the ${\rm FRO} \rightarrow {\rm QFO}$ phase transition [see for instance the curves labeled as $d = -1.4$, $-1.45$ in Fig.~\ref{fig5}(a) and the curves~$\Delta = 3.3$, $3.5$ in Fig.~\ref{fig5}(b)]. Moreover, several interesting regions with reentrant phase transitions can also be observed in phase diagrams of the diamond-like decorated square lattice. More specifically, the square lattice shows the reentrant behaviour with either two or three consecutive critical points [see the curves $d = -1.45, -1.47$ in Fig.~\ref{fig5}(c) and the curve $\Delta = 3.5$ in Fig.~\ref{fig5}(d)]. According to the ground-state analysis, this lattice exhibits two or three consecutive phase transitions of second order only if the anisotropy parameter~$\Delta$ and the AZFS parameter $d$ are selected sufficiently close to the phase transitions FRO--FRU and QFO--FRU. In general, the origin of this non-trivial phenomenon lies in a mutual competition between the easy-plane ($D < 0$, $\Delta > 1$) and easy-axis ($D > 0$, $\Delta < 1$) interactions, while the latter one is also supported by the Ising interaction $J_{\rm I}$. Evidently, the reentrance occurs either if the AZFS parameter $d$ takes the negative values and the exchange anisotropy $\Delta$ is of an easy-axis type or if the reverse case is considered [see the curves $d = -1.45, -1.47$ and also the curve $\Delta = 3.5$ in Figs.~\ref{fig5}(c) and (d), respectively]. By comparing the formerly published results for the diamond-like decorated triangular lattice (see Ref.~\cite{Jas08}) with the results shown in Fig.~\ref{fig5}, one may conclude that the parameter region with reentrant phase transions enlarges with the increasing coordination number of the lattice. Indeed, the mixed spin-$1/2$ and spin-$1$ Ising-Heisenberg model on the decorated triangular lattice exhibits two or three critical points also for $D < 0$ and $\Delta > 1$. Besides, this lattice as the only one is characterised by very interesting dependencies of $t_c$ that have a shape of semi-closed and closed loops (see Fig.~2 in Ref.~\cite{Jas08}). 
\begin{figure}[htb]%
\vspace*{-0.25cm}
\includegraphics*[width=7.5cm]{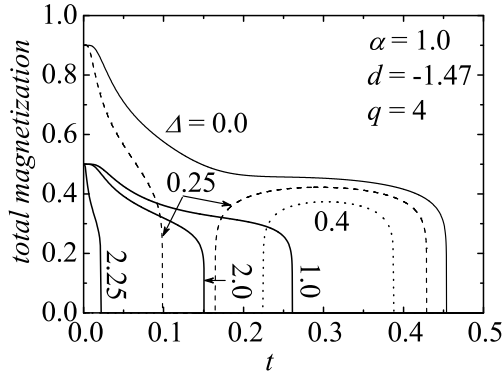}%
\vspace*{-0.5cm}
\caption{Temperature dependencies of the total magnetisation of the decorated square lattice with the interaction ratio
         $\alpha = 1.0$ and the AZFS parameter $d = -1.47$ for several values of the exchange anisotropy $\Delta$.}
\label{fig6}
\end{figure}
In these parts of phase diagrams, the system starts from the disordered ground state before entering the spontaneously ordered phase (either FRO or QFO) at lower critical temperature $t_{c1}$, which very quickly disappears due to strong thermal fluctuations at upper critical temperature $t_{c2}$.

The results displayed in Fig.~\ref{fig5} can be convincingly evidenced by thermal dependencies of the total magnetisation, which represents the order parameter for both ferromagnetic phases FRO and QFO. For this purpose, some temperature variations of the total magnetisation $m = (m_{\rm i}^z + 4 m_{\rm h}^z)/5$ of the diamond-like decorated square lattice are displayed in Fig.~\ref{fig6} for the interaction ratio~$\alpha = 1.0$, the AZFS parameter $d = -1.47$ and several values of the exchange anisotropy $\Delta$. Evidently, the $m(t)$ curve exhibits reentrant transitions with two or three consecutive critical points for $\Delta = 0.4$ and $0.25$, respectively, in accordance with the finite-temperature phase diagram displayed in Fig.~\ref{fig4}(c). Note that both these values of the exchange anisotropy relate to the regions rather close to the phase transition between the ferromagnetically ordered phase and the disordered FRU phase. On the other hand, when one is considering the exchange anisotropies that are sufficiently far from the phase transitions FRO--FRU and QFO--FRU, then the plotted magnetisation curves $m(t)$ exhibit R-type dependencies with a single critical temperature (see e.g.~the curves $\Delta = 0.0$ and $1.0$). According to the ground-state analysis, the initial values $m = 0.9$ and~$0.5$ correspond to the FRO and QFO phases, respectively [for clarity see also the set of Eqs.~(\ref{eq:FRO}) and~(\ref{eq:QFO})].

\begin{figure*}[htb]%
\vspace*{-0.75cm}
\begin{center}
\includegraphics*[width=14.5cm]{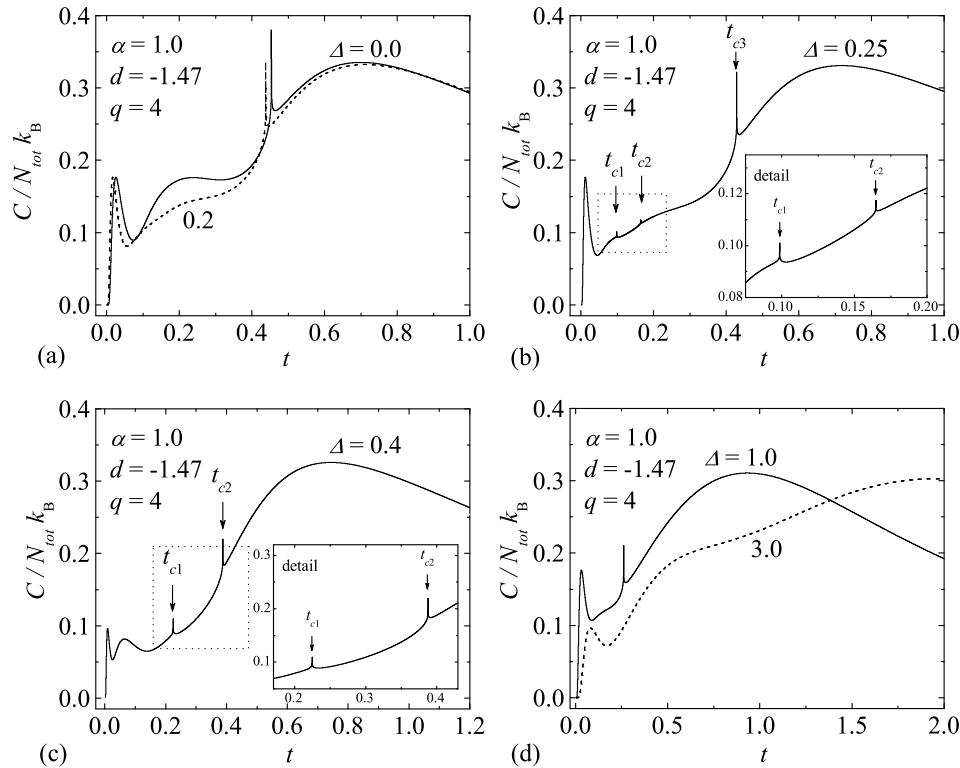}
\end{center}
\vspace*{-0.75cm}
\caption{Temperature dependencies of the specific heat for the diamond-like decorated square lattice with the
         interaction ratio $\alpha = 1.0$, the AZFS parameter $d = -1.47$ and several values of the exchange anisotropy
         $\Delta$. Insets in Figs.~\ref{fig7}(b) and (c) afford details from regions with logarithmic singularities,
         which would not be clearly visible within the displayed scale in these figures.}
\label{fig7}
\end{figure*}

Finally, let us close our discussion by exploring temperature dependencies of the specific heat. For illustation, some typical thermal variations of the specific heat of the Ising-Heisenberg model on the diamond-like decorated square lattice are plotted in Fig.~\ref{fig7}. To enable a direct comparison, we have chosen the values of the interaction ratio $\alpha$, the AZFS parameter $d$ and the exchange anisotropy parameter $\Delta$ so as to match all temperature dependencies of the total magnetisation plotted in Fig.~\ref{fig6}. In this manner, the displayed set of specific heat curves reflects a comprehensive picture of the finite-temperature behaviour of the investigated spin system. If the exchange anisotropy parameter $\Delta$ is chosen to be sufficiently far from the phase transitions FRO--FRU and QFO--FRU, then temperature dependencies of the specific heat with just one logarithmic singularity are observable at the continuous (second-order) phase transition between the spontaneously ordered and disordered phases [see Figs.~\ref{fig7}(a) and (d)]. It should be also mentioned that these thermal dependencies exhibit yet another two or three round Schottky-type maxima in addition to a pronounced logarithmic divergence. On the other hand, remarkable specific heat curves with two or three logarithmic singularities can also be detected when the value of $\Delta$ is taken from the region close to the phase transitions FRO--FRU and QFO--FRU [see Figs.~\ref{fig7}(b) and (c)]. Among other matters, the observed logarithmic singularities provide an independent check of the reentrant behaviour previously discussed by finite-temperature phase diagrams, as well as, temperature dependencies of the total magnetisation. Besides, one or two local maxima can also be detected in low-temperature tails of all displayed specific heat curves in addition to a broad Schottky-type maximum to emerge at high temperatures. It is quite reasonable to expect that the origin of observed low-temperature maxima lies in strong thermal excitations to a spin configuration rather close in energy to the ground state.

\section{Concluding remarks}
\label{Sec4}

In the present paper, the ground-state and critical behaviour of the mixed spin-$1/2$ and spin-$1$ Ising-Heisenberg model on diamond-like decorated honeycomb and square lattices has been investigated within the framework of the generalised decoration-iteration mapping transformation. Using this rigorous procedure, the exact solution for the investigated mixed-spin model has been obtained by establishing a precise mapping equivalence with the spin-$1/2$ Ising model on the corresponding undecorated planar lattice with the known exact solution.

The main emphasis of this work was to bring a deeper insight into the effect of quantum fluctuations and geometric frustration generated by the mutual competition between the Heisenberg interaction, Ising interaction and AZFS parameter on magnetic properties of the planar mixed-spin model with diamond-like decorations at zero as well as non-zero temperatures. Our further goal was to shed light on how the critical behaviour of this system depends on its lattice topology. The obtained results clearly demonstrate that an interplay between the Ising interaction favouring the spin arrangement along the $z$-axis and the easy-plane XXZ Heisenberg interaction favouring the short-range ferromagnetic order in the $xy$ plane gives rise to two interesting ground states (QFO and FRU) with entangled states of the Heisenberg spins. 
Moreover, the appearance of one of those phases (namely, the appearance of the FRU phase) verified the known fact that the competition between the ferromagnetic Ising and ferromagnetic easy-plane Heisenberg interactions leads to occurrence of the short-range antiferromagnetic spin order along the $z$-axis, which is accompanied by the short-range ferromagnetic spin order in the $xy$ plane~\cite{Str02,Str02a,Str06}. Hence, one may conclude that this behaviour represents a general feature of all quantum models with a mixture of easy-axis and easy-plane bonds, since it can be found regardless of the lattice topology as well as the spatial dimensionality of the investigated model system. 

The most interesting result to emerge from the present study consists in the exact evidence of the existence of an reentrant phase transitions with either two or three consecutive critical points. The existence of this non-trivial phenomenon has also been evidenced by temperature dependencies of the total magnetisation, as well as, remarkable thermal variations of the specific heat exhibiting two or three logarithmic singularities. As it has been proved, the higher is the coordination number of the lattice, the more pronounced the reentrance is and more diverse critical behaviour of the system may be found.

Finally, it is worthwhile to remark that even though our theoretical investigation of the mixed spin-$1/2$ and spin-$1$ Ising-Heisenberg model on diamond-like decorated planar lattices was mainly aimed at providing a deeper insight into cooperative and quantum features of this exactly solvable model, we hope that our results might stimulate researches on possible experimental realizations of this interesting quantum spin model and confirm our theoretical predictions. From this perspective, the most promising approach in experimental realizations of our model system may represent a targeted design of cyano-based polymeric coordination compounds or their isostructural analogues. For instance, a series of bimetallic polymeric coordination compounds {Cu(L)}$_3$ [Fe(CN)$_6$](ClO$_4$)$_2$$\cdot$nH$_2$O, L = N-(3-aminopropyl)-1,3-propanediamine~\cite{Zha00,Hon04} or N-(2-aminoethyl)-1,3-propanediamine~\cite{Tra01}, with the diamond-like decorated honeycomb network structure may represent useful starting point for this rational synthesis. In this series, the divalent Cu$^{\rm II}$ and Fe$^{\rm II}$ metal atoms reside decorating and nodal sites of the diamond-like decorated honeycomb lattice, respectively [Fig.~\ref{fig1}(a)]. Unfortunately, the divalent Fe$^{\rm II}$ atoms are due to a strong ligand field of the cyano group in the diamagnetic low-spin state with $S=0$. Similarly, the bimetallic polymeric coordination compounds [W{(CN)$_4$ Fe (H$_2$O)$_2$}$_2$]$\cdot$nH$_2$O~\cite{Pil01} and [W{(CN)$_4$ Co(H$_2$O)$_2$}$_2$]$\cdot$nH$_2$O~\cite{Her03} with the diamond-like decorated square network structure [Fig.~\ref{fig1}(b)] have also been reported quite recently, but here, the tetravalent W$^{\rm IV}$ metal atoms residing nodal sites of the diamond-like decorated square lattice are due to a strong ligand field of the cyano group in the diamagnetic state $S=0$.

\begin{acknowledgement}
This work was financially supported by the Slovak Research and Development Agency under the contract LPP-0107-06 and by Ministry of Education of SR under the grant No.~VEGA 1/0128/08.
\end{acknowledgement}


\begin{thebibliography}{[100]}

\bibitem{Lhu02}%
C.~Lhuillier and G.~Misguich, in: High Magnetic Fields: Applications in Condensed Matter Physics and Spectroscopy, edited by C.~Berthier, L.\,P.~L\'evy, and G.~Martinez, Lect. Notes Phys. (Springer-Verlag, Berlin, 2002), vol.~595.

\bibitem{Rich04}%
J.~Richter, J.~Schulenburg, and A.~Honecker, in: Quantum Magnetism, edited by U.~Schollw\"ock, J.~Richter, D.\,J.\,J.~Farnell, and R.\,F.~Bishop, Lect. Notes Phys. (Springer-Verlag, Berlin, 2004), vol.~645.

\bibitem{Mis04}%
G.~Misguich and C.~Lhuillier, in: Frustrated Spin Systems, edited by H.\,T.~Diep (World Scientific, Singapore, 2004), chap.~5.

\bibitem{Vil77}%
J.~Villian, J. Phys. C: Solid State Phys. \textbf{10}, 1717 (1977).

\bibitem{Vil80}%
J.~Villian, R.~Bidaux, J.\,P.~Carton et al., J. Physique \textbf{41}, 1263 (1980).

\bibitem{Zhi03}%
M.\,E.~Zhitomirsky, Phys. Rev. B \textbf{67}, 104421 (2003).

\bibitem{Hon05}%
A.~Honecker and J.~Richter, Condens. Matter Phys. \textbf{8}, 813 (2005).

\bibitem{Der06}%
O.~Derzhko and J.~Richter, Eur. Phys. J. \textbf{52}, 23 (2006).

\bibitem{Sch07}%
J.~Schnack, R.~Schmidt, and J.~Richter, Phys. Rev. B \textbf{76}, 054413 (2007).

\bibitem{Hon04}%
A.~Honecker, J.~Schulenburg, and J.~Richter, J. Phys.: Condens. Matter \textbf{16}, S749 (2004).

\bibitem{Bar82}%
J.\,A.~Barclay, W.\,A.~Steyert, Cryogenics \textbf{22}, 73 (1982).

\bibitem{Lac83}%
A.\,F.~Lacaze, R.~Beranger, G.~Bon-Mardion {\it et al.}, Cryogenics \textbf{23}, 427 (1983).

\bibitem{Ina96}%
T.~Inami, Y.~Ajitro, and T.~Goto, J. Phys. Soc. Jpn. \textbf{65}, 2374 (1996).

\bibitem{Tan02}%
H.~Tanaka, T.~Ono, H.~Aruga Katori {\it et al.}, Prog. Theor. Phys. Suppl. \textbf{145}, 101 (2002).

\bibitem{Ono03}%
T.~Ono, H.~Tanaka, H.~Aruga Katori {\it et al.}, Phys. Rev. B \textbf{67} 104431 (2003).

\bibitem{Ono04}%
T.~Ono, H.~Tanaka, O.~Kolomiyets {\it et al.}, J. Phys.: Condens. Matter \textbf{16}, S773 (2004).

\bibitem{Ono05a}%
T.~Ono, H.~Tanaka, T.~Nakagomi {\it et al.}, J. Phys. Soc. Jpn. \textbf{74}, 135 (2005).

\bibitem{Ono05b}%
T.~Ono, H.~Tanaka, O.~Kolomiyets {\it et al.}, A. Prog. Theor. Phys. Suppl. \textbf{159}, 217 (2005).

\bibitem{Svi03}%
L.\,E.~Svistov, A.\,I.~Smirnov, L.\,A.~Prozorova {\it et al.}, Phys. Rev. B \textbf{67}, 094434 (2003).

\bibitem{Pro03}%
L.\,A.~Prozorova, L.\,E.~Svistov, A.\,I.~Smirnov {\it et al.}, J. Magn. Magn. Mater. \textbf{258--259}, 394 (2003).

\bibitem{Svi06}%
L.\,E.~Svistov, A.\,I.~Smirnov, L.\,A.~Prozorova {\it et al.}, Phys. Rev. B \textbf{74}, 024412 (2006).

\bibitem{Smi07}%
A.\,I.~Smirnov, H.~Yashiro, S.~Kimura {\it et al.}, Phys. Rev. B \textbf{75}, 134412 (2007).

\bibitem{Nar04}%
Y.~Narumi, Z.~Honda, K.~Katsumata {\it et al.}, J. Magn. Magn. Mater. \textbf{272--276}, 878 (2004).

\bibitem{Kag99}%
H.~Kageyama, K.~Yoshimura, R.~Stern {\it et al.}, Phys. Rev. Lett. \textbf{82}, 3168 (1999).

\bibitem{Oni00}%
K.~Onizuka, H.~Kageyama, Y.~Narumi {\it et al.}, J. Phys. Soc. Jpn. \textbf{69}, 1016 (2000).

\bibitem{Kag01}%
H.~Kageyama, Y.~Narumi, K.~Kindo {\it et al.}, J. Alloys Compd. \textbf{317--318}, 177 (2001).

\bibitem{Kag02}%
H.~Kageyama, Y.~Ueda, Y.~Narumi {\it et al.}, Prog. Theor. Phys. Suppl. \textbf{145}, 17 (2002).

\bibitem{Mich06}%
S.~Michimura, A.~Shigekawa, F.~Iga {\it et al.}, Physica B \textbf{378--380}, 596 (2006).

\bibitem{Yos06}%
S.~Yoshii, T.~Yamamoto, M.~Hagiwara {\it et al.}, J. Phys.: Conf. Ser. \textbf{51}, 59 (2006).

\bibitem{Iga07}%
F.~Iga, A.~Shigekawa, Y.~Hasegawa {\it et al.}, J. Magn. Magn. Mater. \textbf{310}, e443 (2007).

\bibitem{Sie08}%
K.~Siemensmeyer, E.~Wulf, H.-J.~Mikesha {\it et~al.}, Phys. Rev. Lett. \textbf{101}, 177201 (2008).

\bibitem{Str02}%
J.~Stre\v{c}ka and M.~Ja\v{s}\v{c}ur, Phys. Status Solidi B \textbf{233}, R12 (2002).

\bibitem{Can04}%
L.~\v{C}anov\'a, J.~Stre\v{c}ka, and M.~Ja\v{s}\v{c}ur, Czech. J. Phys. \textbf{54}, D579 (2004).

\bibitem{Can06}%
L.~\v{C}anov\'a, J.~Stre\v{c}ka, and M.~Ja\v{s}\v{c}ur, J. Phys.: Condens. Matter \textbf{18}, 4967 (2006).

\bibitem{Can08}%
L.~\v{C}anov\'a, J.~Stre\v{c}ka, J.~Dely et al., Acta Phys. Pol. \textbf{113}, 449 (2008).

\bibitem{Jas08}%
M.~Ja\v{s}\v{c}ur, J.~Stre\v{c}ka, and L.~\v{C}anov\'a, Acta Phys. Pol. \textbf{113}, 453 (2008).

\bibitem{Str09}%
J.~Stre\v{c}ka, L.~\v{C}anov\'a, T.~Lu\v{c}ivjansk\'y {\it et al.}, J. Phys.: Conf. Ser. \textbf{145}, 012058 (2009).

\bibitem{Can09}%
L.~\v{C}anov\'a, J.~Stre\v{c}ka, and T.~Lu\v{c}ivjansk\'y, to be published in Condens. Matter Phys. (arxiv:0903.4566)

\bibitem{Fis59}%
M.\,E.~Fisher, Phys. Rev. \textbf{113}, 969 (1959).

\bibitem{Syo72}%
I.~Syozi, in: Phase Transition and Critical Phenomena, edited by C.~Domb and M.\,S.~Green (Academic Press, New York, 1972), vol.~1.

\bibitem{Roj09}%
O.~Rojas, J.\,S.~Valverde, S.\,M.\,de~Sousa, Physica A \textbf{388}, 1419 (2009).

\bibitem{Per08}%
M.\,S.\,S.~Pereira, F.\,A.\,B.\,F.\,de~Moura, and M.\,L.~Lyra, Phys. Rev. B \textbf{77}, 024402 (2008).

\bibitem{Per09}%
M.\,S.\,S.~Pereira, F.\,A.\,B.\,F.\,de~Moura, and M.\,L.~Lyra, Phys. Rev. B \textbf{79} 054427 (2009).

\bibitem{Lis}%
B.\,M.~Lisnii and O.\,V.~Derzhko, private communications.

\bibitem{Rud08a}%
C.~Rudowicz, Physica B \textbf{403}, 1882 (2008).

\bibitem{Rud08b}%
C.~Rudowicz, Physica B \textbf{403}, 2312 (2008).

\bibitem{Bar88}%
J.\,H.~Barry, M.~Khatun, and T.~Tanaka, Phys. Rev. B \textbf{37}, 5193 (1988).

\bibitem{Kha90}%
M.~Khatun, J.\,H.~Barry, and T.~Tanaka, Phys. Rev. B \textbf{42}, 4398 (1990).

\bibitem{Bar91}%
J.\,H.~Barry, T.~Tanaka, M.~Khatun, and C.\,H.~M\'unera, Phys. Rev. B \textbf{44}, 2595 (1991).

\bibitem{Cal63}%
H.\,B.~Callen, Phys. Lett. \textbf{4}, 161 (1963).

\bibitem{Suz65}%
M.~Suzuki, Phys. Lett. \textbf{19}, 267 (1965).

\bibitem{Bal02}%
T.~Balcerzak, J. Magn. Magn. Mater. \textbf{246}, 213 (2002).

\bibitem{Str}%
J.~Stre\v{c}ka, L.~\v{C}anov\'a, in preparation.

\bibitem{Str02a}%
J.~Stre\v{c}ka and M.~Ja\v{s}\v{c}ur, Phys. Rev. B \textbf{66}, 174415 (2002).

\bibitem{Str06}%
J.~Stre\v{c}ka and M.~Ja\v{s}\v{c}ur, Acta Phys. Slovaca \textbf{56}, 65 (2006).

\bibitem{Zha00}%
H.-X. Zhang, Y.-X. Tong, Z.-N. Chen, K.-B. Yu, B.-S. Kang, J. Organomet. Chem. \textbf{598}, 63 (2000).

\bibitem{Hon04}%
Ch.S. Hong, Y.S. You, Inorg. Chim. Acta \textbf{357}, 3271 (2004).

\bibitem{Tra01}%
Z. Tr\'avni\v{c}ek, Z. Sm\'ekal, A. Escuer {\it et al.}, New J. Chem. \textbf{25}, 655 (2001).

\bibitem{Pil01}%
M. Pilkington, M. Gross, P. Franz {\it et al.}, J. Solid St. Chem. \textbf{159}, 262 (2001).

\bibitem{Her03}%
J. M. Herrera, A. Bleuzen, Y. Dromz\'ee {\it et al.}, Inorg. Chem. \textbf{42}, 7052 (2003).

\end{thebibliography}
\end{document}